# LARGE FIELD-OF-VIEW AND MULTI-COLOR IMAGING WITH GaP QUADRATIC METALENSES

Anton V. Baranikov[‡1], Egor Khaidarov[‡1], Emmanuel Lassalle[1], Damien Eschimese[1], Joel Yeo[1,2,3], N. Duane Loh[2,4], Ramon Paniagua-Dominguez[*1], and Arseniy I. Kuznetsov[†1]

[1]Institute of Materials Research and Engineering (IMRE), Agency for Science, Technology and Research (A*STAR), 2 Fusionopolis Way, Innovis # 08-03, Singapore 138634

[2]Department of Physics, National University of Singapore, Singapore 117551

[3]Integrative Sciences and Engineering Programme, NUS Graduate School, National University of Singapore, Singapore 119077

[4]Department of Biological Sciences, National University of Singapore, Singapore 117557

[‡]These authors contributed equally to this work

[*]Corresponding author: Ramon_Paniagua@imre.a-star.edu.sg

[†]Corresponding author: Arseniy_Kuznetsov@imre.a-star.edu.sg

**Abstract**

Metalenses, in order to compete with conventional bulk optics in commercial imaging systems, often require large field of view (FOV) and broadband operation simultaneously. However, strong chromatic and coma aberrations present in common metalens designs have so far limited their widespread use. Stacking of metalenses as one of the possible solutions increases the overall complexity of the optical system and hinders the main benefit of reduced thickness and light weight. To tackle both issues, here we propose a single-layer imaging system utilizing a recently developed class of metalenses providing large field of view. Using it, we demonstrate full-color imaging with a FOV of $\sim 100°$. This approach, empowered by computational imaging techniques, produce high quality images, both in terms of color reproduction and sharpness. Suitable for real-time unpolarized light operation with the standard color filters present in prevalent camera systems, our results might enable a pathway for consumer electronics applications of this emerging technology.

**Teaser**

A single-layer metalens imaging system with large field of view (~ 100°) and full-color imaging is proposed, empowered with computational imaging techniques.

**Introduction**

The ongoing trend towards miniaturization and integration in imaging systems has led to a need for novel optical platforms that can provide aberration-free, compact, and cost-effective lenses. In this context, the emerging field of flat optics is widely considered to be a viable solution to meet these requirements. By exploiting the principles of diffraction, a flat lens can create a phase profile that allows focusing an incident beam with significantly reduced thickness compared to conventional bulk optics [1–3]. Flat lenses typically consist of a surface divided into Fresnel zones, which impart a radial phase distribution that varies between 0 and $2\pi$. Together, these zones comprise a $2\pi$-wrapped phase map of the desired phase profile. Constructive interference from the multiple zones at a desired spatial location can then be achieved, resulting in the focusing of the transmitted light.

One can distinguish two types of flat lenses: conventional diffractive lenses (CDL) [4–6] and metalenses based on the recent development of optical metasurfaces [7–10]. While the former rely on the phase accumulated by light propagation in a material thickness, in a similar way as for bulk optics but with a reduced thickness enabled by the phase wrapping, metalenses employ nanostructured surfaces providing subwavelength phase control, which constitutes a radically different paradigm. In the case of metalenses based on high-refractive index and low-loss dielectric materials, which offer the highest efficiencies, phase control can be achieved via three different mechanisms, which are waveguiding [7,11–13], using the geometric (Pancharatnam-Berry) phase of light [8,14–17] or the phase modulation associated with the excitation of optical resonances [18–21] (the latter mechanism has recently been shown to have a topological origin, associated with the creation of pole-zero pairs in the complex eigenfrequency plane of non-Hermitian systems [22]). Albeit different in their principles of operation, CDL and metalenses are both naturally diffractive optical elements with associated chromatic aberrations, which severely restricts their applications in multicolor imaging. Indeed, the incident light deflection is either governed by the grating equation for CDL or generalized Snell's law for metalenses [23], in both of which the deflection angle has a wavelength dependence. This translates into shorter wavelengths experiencing a longer focal length and vice versa, where the fractional change in the focal length $\Delta f$, that is the axial (or longitudinal) chromatic aberration, is equal to the fractional change in the wavelength $\Delta \lambda$ [24]:

$$\frac{\Delta f}{f} = \frac{\Delta \lambda}{\lambda} \tag{1}$$

with $\lambda$ and $f$ being the nominal (central) wavelength and focal length, respectively.

To address this problem, researchers have developed several approaches to attempt to achieve achromatic flat lenses. An achromatic diffractive lens (ADL) usually operates at higher orders diffraction, providing the same deflection angle for several discrete harmonic wavelengths [25, 26]. Though advanced numerical optimization helped to increase the focusing efficiency [2, 27, 28], a recent study demonstrated considerable limitations on the achievable Fresnel number (FN), which translates into limited numerical apertures (NA) or lens sizes for a given focal length [29]. This issue arises from the residual chromatic aberration in between adjacent harmonic wavelengths. The associated chromatic focal shift cannot be fully eliminated, since it is inversely proportional to the structure depth and refractive index used in the ADL, which is usually limited. Regarding achromatic metalenses (AMLs), their principle leverages on the additional degrees of freedom given by the design of the individual meta-atoms. Typically, AMLs exploit a large meta-atom library to find elements with an appropriate group delay (GD) and group delay dispersion (GDD) to compensate the chromatic aberrations over a certain wavelength range [30–35]. However, similarly to ADL, AMLs have limited NA or lens size for a given focal length, in this case bounded by the achievable GD and GDD values [29, 32,36,37]. Recently, this issue was addressed by compensating the phase within each Fresnel zone and subsequently optimizing the phase discontinuities at the zone boundaries [38]. It is important to note that in virtue of the task complexity, the optimization was done at three discrete wavelengths. As a result, a high-NA (0.7), millimeter-scale metalens focusing at three red-green-blue (RGB) wavelengths was demonstrated.

Despite the tremendous progress made in achromatic flat lenses, these approaches fail to address another crucial imaging characteristic, namely achieving a large field of view (FOV) [29, 38]. In this regard, while it has been shown that metalenses can outperform CDL in terms of angular coverage [39], conventional diffraction-limited metalenses with a hyperbolic phase profile still suffer from strong off-axis aberrations, severely curtailing their FOV [40]. To circumvent this issue, a number of works have recently proposed various solutions based on aplanatic metalenses [41], numerical optimization [42–44], metalens doublets [45, 46] or novel phase profiles, such as the so-called quadratic phase profile [21, 47–49]. Among them, the latter has the advantage of being planar and single-layer and has been shown to provide imaging beyond 100° FOV. Indeed, such metalenses provide a large FOV by having an effective working area (with a diameter of *2f*), which is transversely shifted for different angles of incidence ($\varphi$), according to the formula *f*sin$\varphi$ [49]. In the working area, the phase distribution of the transmitted light is preserved, which results in eliminating all off-axis aberrations. This comes at the cost of having spherical aberrations which result in non-diffraction-limited performances, and thus slightly smaller imaging resolution, though sufficient for most general purposes (quadratic metalenses typically have $NA \sim 0.3 - 0.4$, which is higher than a typical smartphone camera with NA of 0.2 [29]) [21, 48, 49].

Interestingly however, because of these intrinsic spherical aberrations, quadratic metalenses present a focus elongated axially, and thus have a certain depth-of-focus (DOF), given by [50]:

$$\text{DOF} \sim \frac{\lambda}{NA^2} \tag{2}$$

This DOF leads to a certain spectral operation bandwidth. Indeed, qualitatively, the focal length shift due to chromatic aberrations is still acceptable when it does not exceed the DOF of the flat lens, that is as long as it remains within $\Delta f \sim \text{DOF}$. Hence, from Eq. (1), the working spectral bandwidth of a flat-lens can be given by the criterion:

$$\Delta \lambda \sim DOF \times \frac{\lambda}{f} \tag{3}$$

The idea of leveraging or engineering an extended DOF for achromatic operation was considered in [50] and full-color imaging was demonstrated in [51,52] using this concept. However, the limited FOV problem remained unaddressed in those works. The hypothesis for quadratic metalenses to operate within a wide spectral bandwidth was put forth in [47], and just recently started to be explored experimentally in [53], nonetheless, the practical bandwidth ranges or use for multi-color imaging have yet to be fully explored.

In this work, we leverage on extended DOF and wide bandwidth benefits of quadratic metalenses to demonstrate a practical optical system tackling both the broadband multi-color and large FOV challenges simultaneously. Our proposed system is planar and single-layer, with uniform thickness, and thus, suitable for one-step lithography fabrication with high throughput (e.g. photolithography [54] or nanoimprint [55, 56]). It consists of three distinct metalenses working at different color channels, in the red, green and blue (RGB). For multi-color imaging system to be compact, one should engineer the metalenses working bandwidth to match the spectral bandwidth of the color filters of the detectors used, eliminating the need for additional filters. While multi-band imaging realization based on parallel metalenses was suggested in the literature [45, 57], we refine this idea here to a single layer by using quadratic metalenses combined with the color filters present inside a standard camera, to realize a full-color imaging system. The metalens unit elements are circular in cross section, allowing for polarization-independent operation. Moreover, our solution employs gallium phosphide (GaP), a material with high potential for metasurface-based devices operating across the visible [58], as it provides a high-refractive index ($n > 3.3$) and negligible losses in the whole wavelength range. We underline the imaging potential of quadratic metalenses by demonstrating their broadband operation: we experimentally show that point spread function (PSF) and modulation transfer function (MTF) remain virtually unchanged over a continuous 40 nm bandwidth. This result is far beyond what widely used hyperbolic metalenses can achieve. Next, by combining the images formed by three quadratic metalenses working at different RGB channels, we demonstrate large FOV ($\sim 100°$) multi-color imaging, with excellent color reproduction in the CIELAB color reproduction assessment. Finally, in conjunction with computation imaging techniques (Wiener and EigenCWD), we obtain image quality in terms of color reproduction and sharpness among the best demonstrated so far with flat optics.

**Results**

Fig. 1a illustrates the concept of our work. The light coming from an object is focused onto a color charge-coupled device (CCD) camera by three quadratic metalenses, fabricated on the same substrate and having the same thickness. Each metalens is designed for operating in a distinct wavelength channel, namely red (R), green (G) and blue (B) wavelengths, respectively. The CCD camera's inherent internal color filters are employed to generate individual R, G, and B images, which are subsequently merged to yield a full-color image. All our lenses are fabricated on the same substrate and possess an equal focal length ($f$), thus ensuring the same imaging magnification across all RGB channels and facilitating easy post-processing procedures.

To prove the concept, we design and fabricate three quadratic metalenses (R, G and B) operating, respectively, at $\lambda_R = 620$ nm, $\lambda_G = 530$ nm and $\lambda_B = 460$ nm wavelengths. The lenses have the same diameter $D = 200\,\mu$m and focal length $f = 83\,\mu$m. They are realized by encoding a wrapped discretized quadratic phase profile:

$$\Phi_i(r) = \Phi_i(0) - \frac{2\pi}{\lambda_i} \frac{r^2}{2f} \qquad (4)$$

where $i = R, G, B$, using nanopillar waveguides with circular cross-section and the same height $H = 300$ nm, arranged in an hexagonal lattice. Owing to a change in the effective index of the waveguide, the nanopillars impart a variable phase delay as a function of their diameter, and we exploit this mechanism to map the quadratic phase profiles. As a material, we utilize gallium phosphide (*GaP*) in virtue of its negligible absorption over almost the entire visible range (see Supp. Info. Fig. S1), and higher refractive index (*n* = 3.3-3.8) than other suitable transparent materials, such as hafnium oxide (*HfO$_2$*), gallium nitride (*GaN*), titanium dioxide (*TiO$_2$*) and silicon nitride (*Si$_3$N$_4$*) [58]. High refractive index is important to encode a quadratic phase profile at different wavelengths using a uniform pillar height for all metalenses, as it ensures a strong confinement of the optical field inside the waveguides and minimizes parasitic coupling effects that might arise due to the small pitch needed to obtain a large FOV [49]. Note that, in order to have the same *D* and *f* for different $\lambda_i$, the Fresnel number (FN) of the metalenses needs to be adjusted according to $FN_i = \frac{D^2}{4\lambda_i f}$ [29]. In other words, the phase profile to be encoded by the nanopillars is steeper for shorter wavelengths (as can be seen in Eq. (4)). In order to keep a sufficient phase profile sampling and a full phase coverage using the nanopillars elements, we scale the lattice period $p_i$ of each lens according to its operating wavelength $\lambda_i$. In our design, this results in periods $p_R$ = 260 nm, $p_G$ = 220 nm and $p_B$ = 190 nm, for the R, G and B metalenses respectively. Fig. 1b-c show the corresponding finite-difference time-domain (FDTD) calculations of the phase and transmission values for the nanopillar unit cells used to realize the R, G and B metalenses, as a function of the duty cycle, which is the ratio between nanopillar diameters and the hexagonal lattice constant. The *GaP* metalenses are patterned on an *SiO$_2$* substrate using electron beam lithography (EBL) followed by inductively coupled plasma reactive ion etching (ICP-RIE). For more details on the design and fabrication, see Methods. Fig. 1d shows the optical microscope (in false colors) and corresponding SEM images of the fabricated metalenses.

The initial imaging performance of quadratic metalenses can be evaluated from their point spread function (PSF) across the FOV and bandwidth of interest. As an example, we show in Fig. 2 the optical characterization of the R channel metalens ($\lambda_R$ = 620 nm). Therein, Fig. 2a shows a schematic of the experiment and Fig. 2b the experimentally measured PSF. The metalens is illuminated with a collimated laser beam, centered at $\lambda_R$ = 620 nm, with incident angle variation $\varphi$ = 0°, 30°, 50° and laser bandwidth variation $\Delta\lambda$ = 10 nm to 40 nm. From Fig. 2b, we can conclude that PSF is almost insensitive to the bandwidth for $\varphi = 0°$. This confirms the robustness of quadratic metalenses against axial chromatic aberrations, coming from their intrinsic spherical aberrations, which result in an axial elongation of the focal spot (or, in other words, a certain DOF). For higher angles of incidence $\varphi = 30°$ and $\varphi = 50°$, the PSF slightly degrades. The observed broadening is in excellent agreement with simulations (see Supp. Info. Fig. S2 b), and is related to the lateral (or transversal) chromatic aberrations of the metalens (details in Supp. Info. Section 1). Rigorous MTF analysis for R,G,B channel metalenses is given in Supp. Info. Section 2.

To further evaluate the robustness of the quadratic metalenses in terms of bandwidths and angles of incidence, initially revealed by the PSF analysis, we conducted an imaging experiment of a standard USAF 1951 resolution test target (element 2, group -2 with a spatial frequency of 0.28 cycles/mm). As a benchmark for performance comparison, we consider the widely used, hyperbolic phase profile metalens. Fig. 3a shows monochromatic imaging simulations for hyperbolic (top panel) and quadratic (bottom panel) phase profile metalenses for $\varphi = 0°$. The designed central wavelength and diameter are $\lambda_R = 620$ nm and 200 $\mu$m for both lenses. The focal length of the quadratic metalens is 83 $\mu$m, while for the hyperbolic lens – 173 $\mu$m to match the effective $NA = 0.5$ of the quadratic metalens. The image produced by the hyperbolic lens becomes significantly blurred for bandwidths larger than $\Delta\lambda = 10$ nm, in contrast to the quadratic metalens, which exhibits virtually unchanged imaging within a bandwidth of $\Delta\lambda = 40$ nm. These simulations are corroborated by experimental measurements with setup schematics given in Fig. 3b. The target element was illuminated by a diffused laser light and the image produced by the R metalens was captured by a CMOS camera (a more detailed schematic of the setup is shown in Supp. Info. Fig. S6). Experimental target element images as a function of incidence angle and bandwidth are given in Fig. 3c. Due to the metalens demagnification, the produced element image for larger angles $\varphi$ is squeezed. One can see that the target element is well-resolved at $\varphi = 0°$ and $\varphi = 30°$ for all the bandwidths considered; for larger angles $\varphi = 50°$ the image partially degraded, except for narrowest bandwidth $\Delta\lambda = 10$ nm where the target is still resolved. In Supp. Info. Section 3 we correlate the MTF obtained from PSF with MTF measured from USAF 1951 target.

We have experimentally demonstrated a relatively broad operational bandwidth ($\sim 40$ nm) of the designed quadratic metalenses over a large FOV (up to 100°). This result is consistent with theoretical estimates: considering an NA $\sim 0.35$ and by combining Eqs. (2) and (3), we expect to have a bandwidth of $\Delta\lambda \sim 38$ nm, 28 nm and 21 nm, for the R, G and B metalenses, respectively. For high-quality imaging, these metalens bandwidths have to match the spectral R, G and B filters of the color camera used, which have an average bandwidth of $\sim 100$ nm if one considers the full width at half maximum (precise quantum efficiencies are given in the Supp. Info. Fig. S8). Even though the bandwidths of our quadratic metalenses here seem to only partially cover the filter spectral bandwidths, next we show that it is sufficient to produce high-quality imaging.

In order to quantify the color imaging performance of the system illustrated in Fig. 1a, we use a standard ColorChecker test chart with 24 painted patches (often referred to as Macbeth chart), depicted in Fig. 4a. To provide uniform object illumination, we utilize a smartphone screen (Xiaomi Redmi Note 8 Pro), which mimics a white color source by three RGB wide bands which nearly match the metalens bandwidths (the white color emission spectrum of the source is shown in Supp. Info. Fig. S9). Fig. 4b displays the composite RGB image obtained as well as individual R, G and B channel components, in the imaging configuration with FOV of 30° x 20°. Note that the merging process also includes a normalization procedure to account for minimum and maximum intensity values (color balance) in each channel. In the resulting RGB image, one can accurately recognize colors of the original for all patches. The details of precise color error calculations are in the Supp. Info. Section 4 and Figure S10. Importantly, the errors vary in the range between 5 and 23 for different patches (CIELAB metric), this result is equivalent to or even better than commercial devices such as iPhone 5s, iPhone 7 or Samsung S10 operating in auto mode [59]. "Veiling glare" or reduced sharpness of the merged image originates from the quadratic metalens imaging behavior and can be significantly improved with the use of deconvolution techniques based on the knowledge of the PSF, as we demonstrate later. Additionally, the ColorChecker was imaged in a configuration with a maximum target FOV of 100° x 67° (Fig. 4c) by bringing the smartphone screen closer to the metalenses. It is noteworthy that for this considerably wide FOV one can still clearly distinguish the colors. The fact that image brightness is reduced towards the periphery, deteriorates the color reproduction (Supp. Info. Fig. S10b). This limitation can be effectively addressed through the intensity correction procedure. For the intensity correction procedure in each R, G and B channel, we characterize the angular efficiency and use the result as a calibration curve (details in Methods and Fig. S11). After this intensity calibration, the overall color error is reduced (see Supp. Info. Fig. S10c). The improvement in color reproduction is more substantial on the periphery, with color errors decreased by 30-40 % in some cases (blue-green, yellow-green and white).

Finally, as a proof-of-use for practical cases, we demonstrate large FOV RGB imaging using a multi-color picture of a still-life genre (Fig. 5a). Again, the image was replicated on a smartphone screen and positioned at two different distances from the metalenses, resulting in FOVs of 50° x 35° (Fig. 5b) and 100° x 67° (Fig. 5c). Even without any post-processing of the obtained RGB images (Fig. 5b-c) the details are still resolved and easily recognized due to the preservation of the MTF at higher spatial frequencies, albeit at a reduced level. To further enhance the image quality, we employed additional computational imaging techniques, including the widely used Wiener filtering, as well as a novel deconvolution algorithm based on the PSF knowledge of the quadratic metalenses (see Supp. Info. Section S5). The images reconstructed by the two techniques (denoted as Wiener filter and EigenCWD, respectively, in Fig. 5b-c) demonstrate greatly improved sharpness and good color reproduction, with better results obtained with our deconvolution algorithm. Indeed, the Wiener filtering presents more elevated noise, a feature that is inevitable for a high-pass filtering method. In contrast, the EigenCWD algorithm has significantly reduced noise levels, although with minor rippling artefacts in the 100° × 67° FOV case, due to the parasitic background noise. To further mitigate these artefacts and improve the image quality it is possible to include an aperture in front of the metalens to suppress background light, as demonstrated in Fig. S5.

For practical applications, it is crucial to consider the computational demands of image processing techniques such as filters or deconvolution algorithms. A fast processing time, low power consumption, and limited computational resources are important factors for portable and compact devices. In this aspect, Wiener filtering presents an advantage with its fast processing time of ∼ 0.3s for a full RGB image (1K x 1K resolution). However, other filters, such as our developed algorithm that belongs to the category of total variation regularizers [51], can result in improved images with better precision and color reproduction, which comes at a cost of increased computational time (up to ∼ 1h for full-color 1.6K x 1.2K pixel image). The application of advanced artificial intelligence techniques could further increase the processing speed of these algorithms.

**Discussion**

In conclusion, we have proposed the use of quadratic metalenses on a single chip for multi-color RGB imaging with a large FOV. The system was designed and fabricated using GaP platform, which is highly suitable for visible range applications and requires only a single layer of lithography process for its manufacturing, making it highly scalable and cost-effective. The system takes advantage of the internal filters present in a CCD camera, reducing the number of components needed and simplifying implementation. Through PSF analysis and imaging experiments, it was demonstrated that the quadratic metalenses provide large FOV imaging across a relatively broad spectral bandwidth of ∼ 40 nm for individual color channels, attributed to its extended DOF. By using a source with a spectrum matching the metalens working bandwidth, we demonstrate RGB imaging with a high-quality color reproduction up to 100° FOV and well-resolved details even for the raw camera image. Further improvements were made to the details, sharpness, and overall color reproduction through the application of a Wiener filter and our own developed deconvolution algorithm. The concept of the metalens system can be further extended from RGB to multi-spectral applications by using a larger array of metalenses, each operating within a certain bandwidth. This would increase the coverage of camera color filters and provide even more accurate color reproduction. Our system could also benefit from incorporating industrial-grade cameras with an advanced color response and balance.

When scaling up the solution to larger metalens sizes, it is important to keep in mind a few key points. While the DOF of quadratic metalenses remains identical for the same NA (Eq. 2), the working bandwidth is reduced with increasing focal length by virtue of Eq. 3. Moreover, the lateral chromatic aberrations, which scale linearly with the focal length (refer to Eq. S3), may further distort the PSF for larger metalenses. However, as demonstrated in the Supp. Info. Section S6, these aberrations can be reduced by incorporating a dispersion-engineering technique, thus enabling the practical application of our system solution to millimeter-diameter metalens systems.

Despite some flaws of quadratic metalenses, Engelberg et.al. [21] previously showed that the efficiency is still sufficient for outdoor light imaging with an acceptable resolution. Here, we leveraged on the extremely high FOV, long depth of focus and broad bandwidth of those and demonstrated a unique solution for simultaneous large FOV and multi-color imaging in real-life scenarios. Moreover, the system is ultracompact, employing a single functional metalenses layer, and suitable for large-scale and high-throughput industrial fabrication. Thus, we believe that our results represent a significant step towards the widespread use of flat optics for low-cost, compact, multi-color, large FOV imaging systems, with numerous applications in consumer electronics and beyond.

## Materials and Methods

### Design

The transmission and phase-delay are computed - by means of Lumerical FDTD software - for periodic arrays of GaP nanopillars with identical diameters, arranged in an hexagonal lattice with lattice constant $p_i$ = 260, 220, 190 nm (for $i$ = R, G, B, respectively), by sweeping over the diameters. For that, a single unit-cell is simulated with periodic boundary conditions in the transversal directions and perfectly matched layers in the longitudinal one. The nanopillars are lying on a $SiO_2$ substrate of refractive index $n$ = 1.46, and their height is set to $H$ = 300 nm in all cases. A monochromatic plane wave with wavelength $\lambda_i$ = 620, 530, 460 nm and amplitude equal to unity is simulated coming from the glass substrate and at normal incidence.

### Sample fabrication

Commercially available GaP on $SiO_2$ on sapphire wafers were purchased from Yangzhou Changelight Co. Initial thickness of GaP was thinned down to match the design using the inductively coupled plasma-reactive ion etching (ICP-RIE, Oxford Plasmalab 100) with $Cl_2$ and $N_2$ gases. Hydrogen silsesquioxane (HSQ, Dow Corning XR-1541) resist followed by current spreading Espacer 300AX01 (Showa Denko) were spin coated on the sample for the electron beam lithography (EBL, Elionix ELS-7000, 100 kV) exposure. All designed RGB metalenses were exposed in one EBL round on the same sample. The resist was developed using tetramethylammonium hydroxide (TMAH) and resulting mask was used to pattern GaP using the ICP-RIE (Oxford Plasmalab 100) at 200°C with $Cl_2$ and $N_2$ gases. The residual HSQ resist on top of the etched structures was preserved, as the hydrofluoric acid (HF) solution commonly used to lift-off the HSQ will damage the $SiO_2$ substrate.

### Optical measurements

The MTF characterization (Supplementary Information, Figure S4) started with the corresponding PSF measurements (Figures 2b). For this, metalenses were illuminated with a collimated tunable laser source (supercontinuum fiber laser SuperK EXTREME equipped with a tunable single line filter SuperK VARIA), which was placed on a rotation arm. The produced PSF was transferred to a CMOS camera (CS895MU Thorlabs) by a homemade optical microscope (100× Olympus plan apo objective with NA = 0.95 and a tube lens with 150 mm focal length). The optical setup is depicted in Supplementary Information, Figure S3. The final MTF plot was produced by a horizontal slice (along the transversal PSF shift upon the increase of $\varphi$, x-axis) of PSF 2D Fourier transform.

In the same way, MTF simulation (Supplementary Information, Figure S2) was performed by PSF Fourier transform. Simulated PSF were obtained by a developed Fourier propagator, which is described elsewhere [49]. Polychromatic PSF was computed by summing up the monochromatic PSF with the wavelength step of 2 nm.

For the USAF 1951 target imaging (Figure 3), the same laser source was utilized. A rotating diffuser was placed in the optical path to generate spatially incoherent light. The image produced by the metalens was transferred to a CMOS camera (Thorlabs CS895MU) by a homemade optical microscope (100× Olympus plan apo objective with NA = 0.95 and a tube lens with 50 mm focal length). The optical setup is depicted in Supplementary Information, Figure S6.

RGB imaging (Figure 4 and Figure 5) was performed using the same optical setup with color CMOS camera (Thorlabs CS895CU). The validation of the color reproduction was done by CIELAB metric. To convert RGB intensity values to *L\*a\*b\** three-dimensional space, a standard software (ImageJ) was utilized.

The focusing efficiency characterization was performed by the same PSF data, measured for the MTF analysis. PSF intensity for all angles of incidence $\varphi$ and bandwidth $\Delta\lambda$ were integrated within a pinhole with a fixed diameter of 6 $\mu$m to account for the spot broadening. Then, the result was divided by a reference PSF intensity. As the reference lens, an antireflective achromatic doublet (Thorlabs AC254-030-AB-ML) was used. To be consistent with our previous publication [49], the final efficiency was scaled to be the ratio between the focusing energy and the light energy transmitted through the effective working lens area (diameter of 2*f*).

The Wiener filter is a deconvolution technique based on the assumption of linear shift-invariant image formation. The ideal image *i(x, y)* is blurred by *PSF (x, y)* of an optical system and corrupted by uncorrelated noise *n(x, y)*. The resulting real image reads: *r(x, y) = s(x, y) + n(x, y)*, where *s(x, y) = i(x, y) \* PSF (x, y)* is blurred image in the absence of the noise. To retrieve *i(x, y)*, one needs to deconvolve measured *r(x, y)* in the Fourier domain applying a proper filtering function to account for noise. This can be written as: $I(u,v) = \frac{R(u,v)F(u,v)}{FPSF(u,v)}$, where *I(u, v)*, *R(u, v)* and *FPSF (u, v)* are Fourier transforms of *i(x, y)*, *r(x, y)* and *PSF (x, y)*, respectively. *F (u, v)* is the applied filtering function. Minimizing $\sum_{u,v}(I(u,v) - \frac{S(u,v)}{FPSF(u,v)})^2$ one ends up with final formula in Fourier domain:

$$I(u,v) = \frac{R(u,v)}{FPSF(u,v)\left(1 + \left|\frac{N(u,v)}{S(u,v)}\right|^2\right)} \qquad (4)$$

where, $N(u, v)$ and $S(u, v)$ are Fourier transforms of the noise $n(x, y)$ and blurred image $s(x, y)$, respectively. Applying the Wiener filter in Figures 4 and 5, the power spectrum ratio $\left|\frac{N(u,v)}{S(u,v)}\right|^2$ was heuristically optimized. The Wiener filtering per channel image took ~100 ms in Figures 4b and 5b (1020x1000 pixels), ~300 ms in Figures 4c and 5c with larger FOV (1616x1160 pixels).

**Acknowledgements**

The authors would like to thank Haizhong Zhang and Leonid Krivitsky (IMRE, A*STAR) for helping with procurement of GaP thin films on $SiO_2$/sapphire substrates.

**Funding**

This work was supported by National Research Foundation of Singapore under Grant No. NRF-NRFI2017-01,

IET A F Harvey Engineering Research Prize 2016,

AME Programmatic Grant No. A18A7b0058 (Singapore),

J.Y. is funded by the A*Star Graduate Scholarship,

J.Y. and N.D.L. thank the IT staff at Centre for Bio-imaging Sciences, National University of Singapore, for their support.

**Author contributions**

A.V.B. performed all the optical measurements, MTF simulation and wrote the first draft. E.K. performed the metalenses nanofabrication, D.E. did the SEM characterization. E.L. designed the metalenses and made the FDTD simulations. J.Y. and N.D.L. developed the deconvolution algorithm for the image reconstruction. A.V.B., R. P-D. and A. I. K. conceived the idea. R. P-D. and A. I. K. supervised the research. All authors analyzed the data and read and corrected the manuscript.

**Competing interest**

The authors declare no competing financial interests.

**Data and materials availability**

All data are available in the main text or the Supplementary Information.

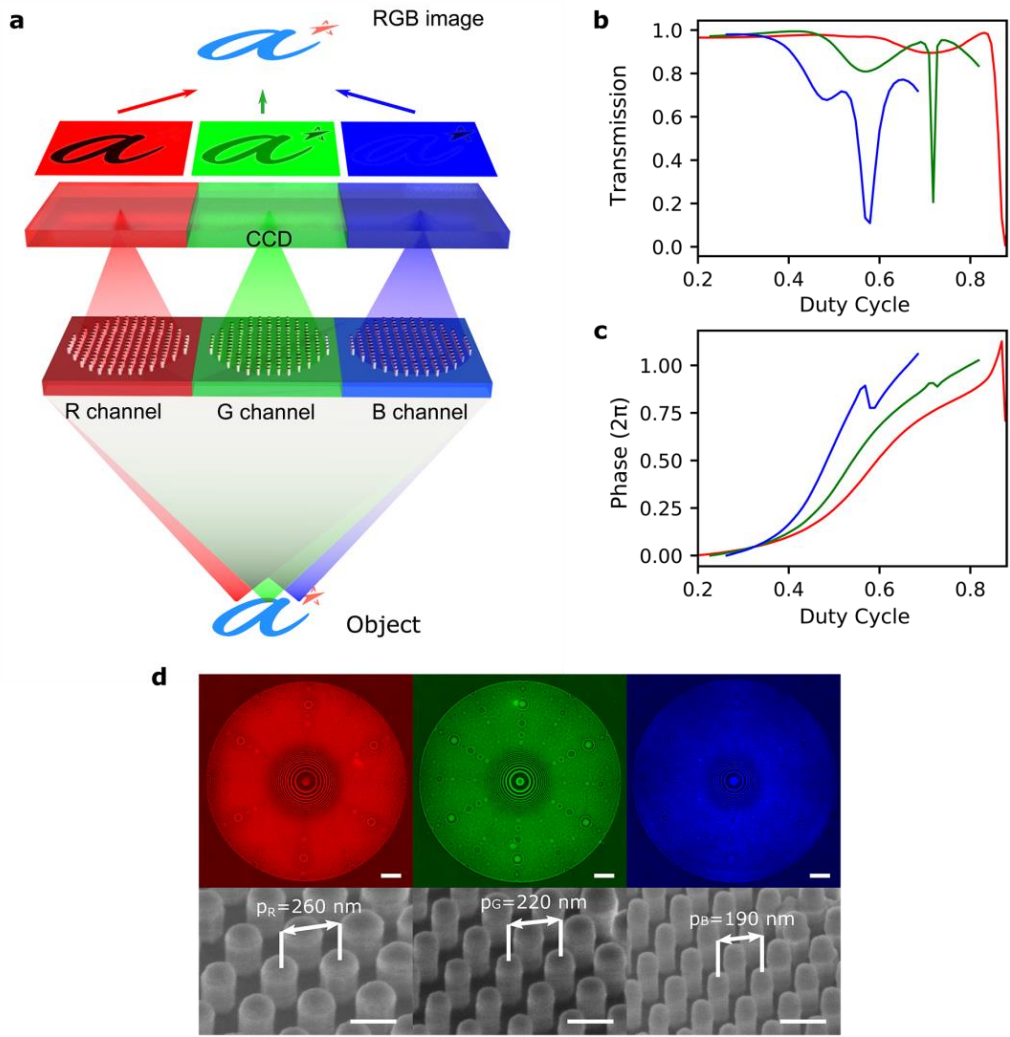

**Figure 1: RGB imaging with large FOV quadratic metalenses. a** An artistic schematic of the device. The light coming from an object is focused on a color CCD camera by three quadratic metalenses on the same chip. Each channel produces R, G and B images which subsequently are merged to produce a final RGB image. **b-c** Simulated phase and transmission values versus duty cycle (ratio between the nanopillar diameters and the lattice constant) for periodic arrays of GaP nanopillars with fixed height of 300 nm used for R,G and B metalenses design (depicted as red, green and blue lines, respectively). **d** Optical microscope (top panel in false colors, the scale bars correspond to 20 $\mu$m) and SEM images (bottom panel, the scale bars correspond to 200 nm) of the fabricated metalenses. The parameters $p_R$ = 260 nm, $p_G$ = 220 nm and $p_B$ = 190 nm denote the designed and fabricated lattice periods.

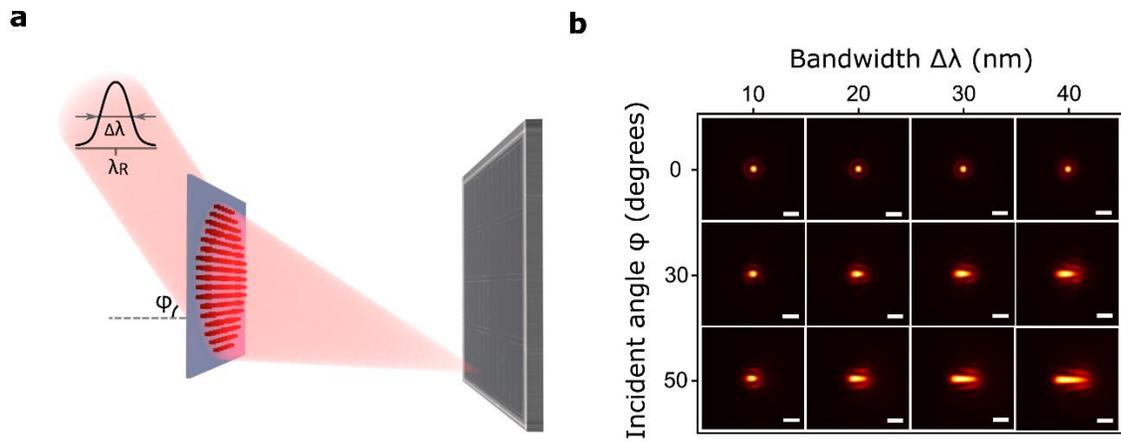

**Figure 2: PSF measurement analysis of the quadratic metalens in red channel. a** Measurement schematics. **b** Measured polychromatic PSF for incident angles $\varphi$ and light source bandwidths $\Delta\lambda$ = 10 nm to 40 nm. The scale bars correspond to 2 $\mu$m.

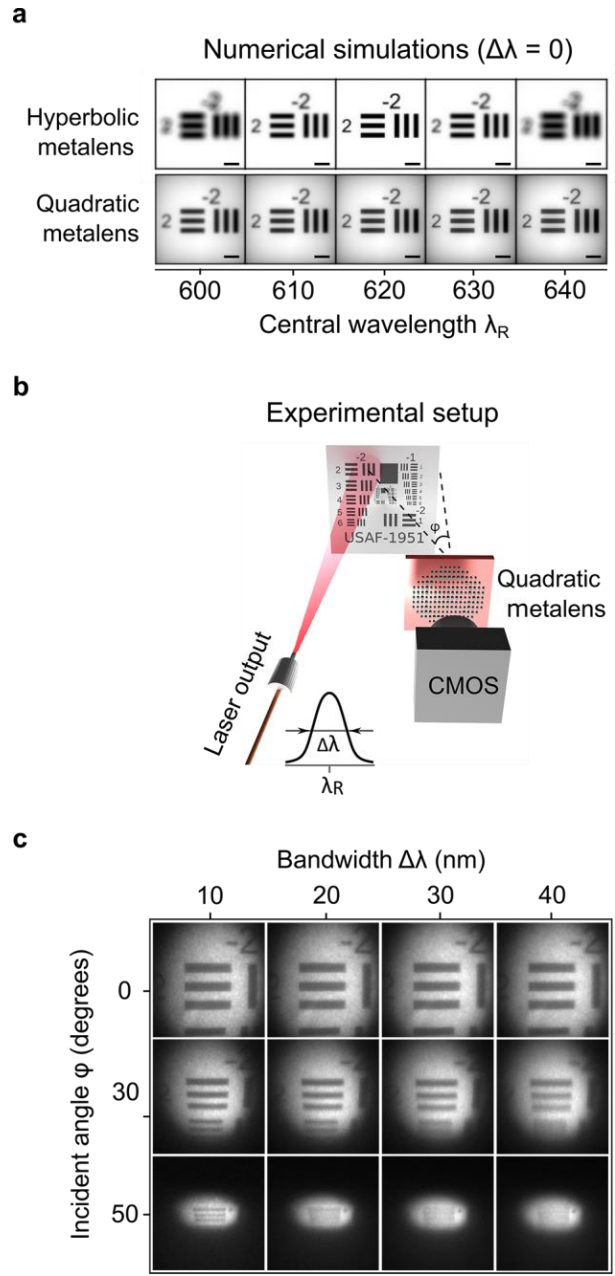

**Figure 3: USAF 1951 test target simulation and characterization with metalenses in red channel. a** The monochromatic imaging simulations for hyperbolic (top panel) and quadratic (bottom panel) phase profile metalenses for $\varphi = 0°$. The designed central wavelength and diameter are 620 nm and 200 $\mu$m for both lenses. The scale bars correspond to 10 $\mu$m for hyperbolic and to 5 $\mu$m for quadratic phase profiles, respectively. **b** Sketch of the optical setup. **c** Imaging with quadratic metalens of the element 2, group -2 for various angles of incidence and source bandwidths.

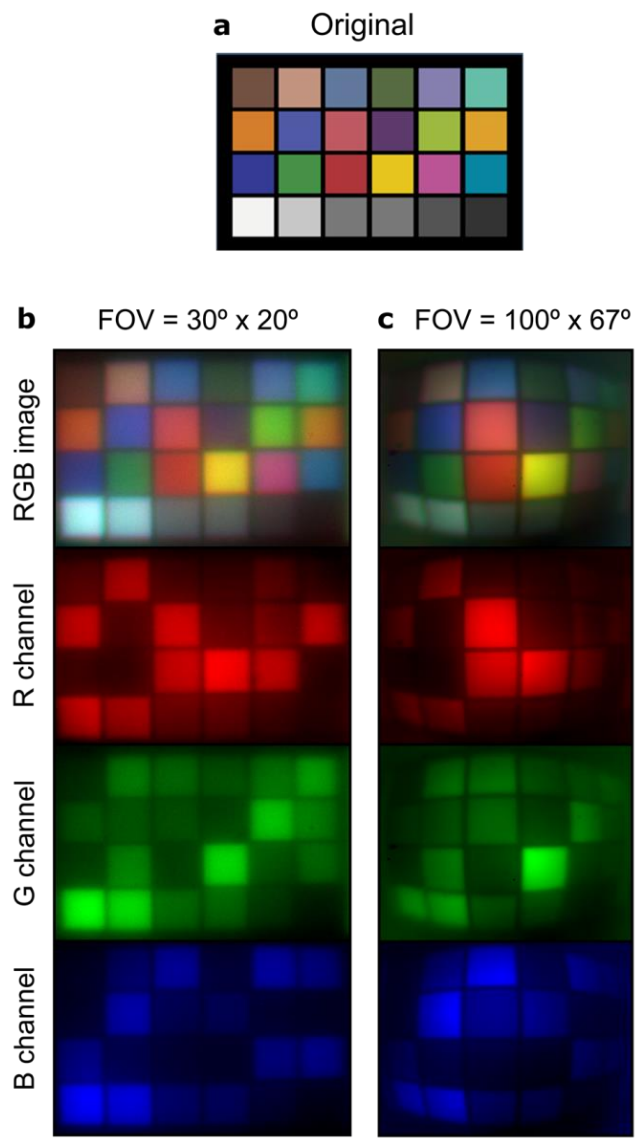

**Figure 4: RGB imaging of ColorChecker chart. a** The original picture. **b,c** The results of RGB imaging for FOV of 30° x 20° and 100° x 67°. R, G and B channels denote raw images, produced by each metalens. RGB image indicates the result of the channels fusion

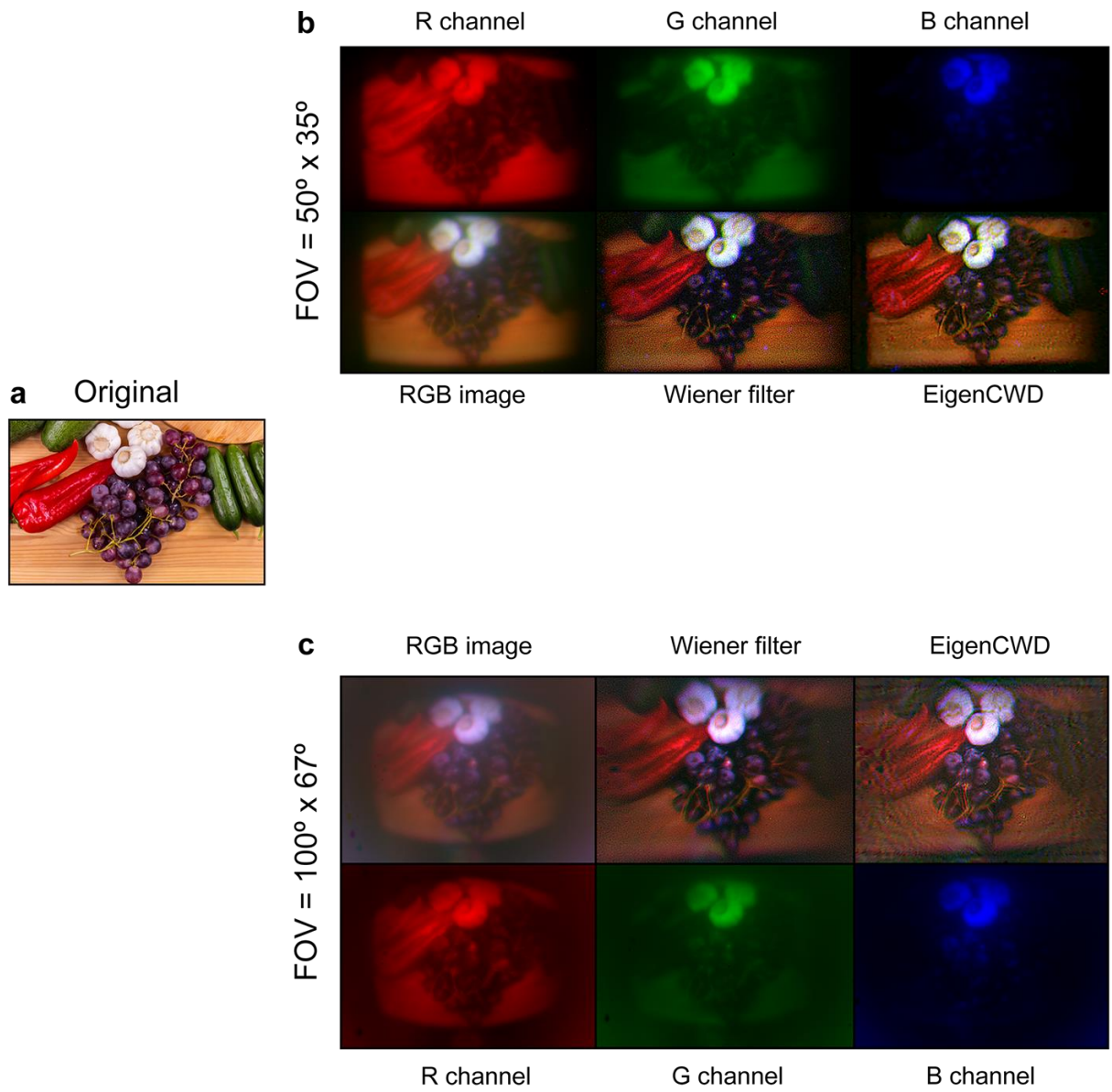

**Figure 5: RGB imaging of a colorful picture. a** The original picture. **b,c** The results of RGB imaging for FOV of 50° x 35° and 100° x 67° respectively. R, G and B channels denote raw images, produced by each metalens. RGB image indicates the result of the channels fusion. Finally, Wiener filter and EigenCWD denote the images reconstructed by applying Wiener filtering and our deconvolution algorithm, respectively.

Supplementary Materials for

# Large Field-of-View and Multi-Color Imaging with GaP Quadratic Metalenses

Anton V. Baranikov[‡1], Egor Khaidarov[‡1], Emmanuel Lassalle[1], Damien Eschimese[1], Joel Yeo[1,2,3], N. Duane Loh[2,4], Ramon Paniagua-Dominguez[*1], and Arseniy I. Kuznetsov[†1]

[1]Institute of Materials Research and Engineering (IMRE), Agency for Science, Technology and Research (A*STAR), 2 Fusionopolis Way, Innovis # 08-03, Singapore 138634

[2]Department of Physics, National University of Singapore, Singapore 117551

[3]Integrative Sciences and Engineering Programme, NUS Graduate School, National University of Singapore, Singapore 119077

[4]Department of Biological Sciences, National University of Singapore, Singapore 117557

[‡]These authors contributed equally to this work

[*]Corresponding author: Ramon_Paniagua@imre.a-star.edu.sg

[†]Corresponding author: Arseniy_Kuznetsov@imre.a-star.edu.sg

**This PDF file includes:**

Supplementary Text (6 sections)

Figs. S1 to S12

Tables S1

References (1 to 11)

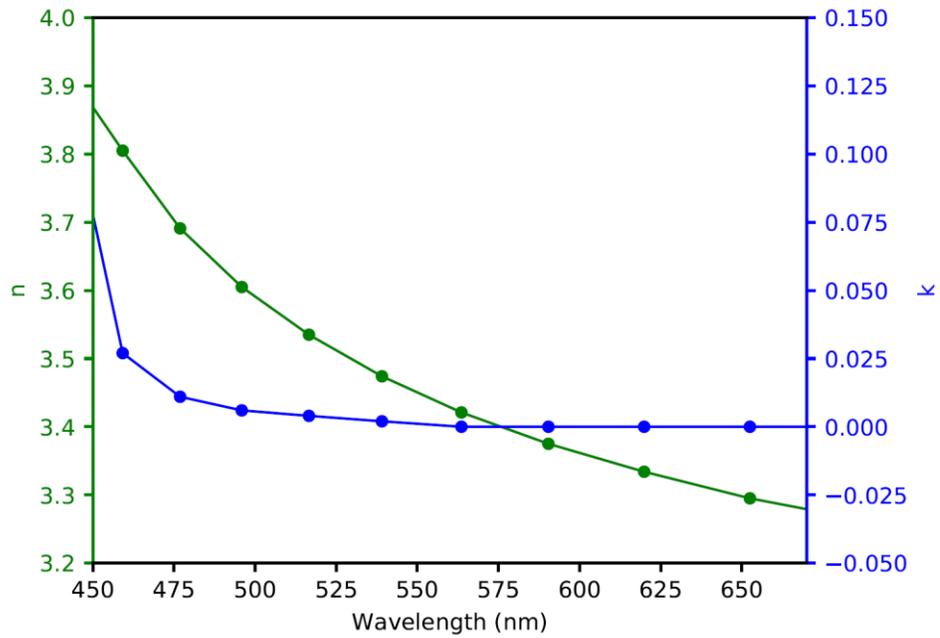

**Figure S1: Measured refractive index of GaP.** Real (n, green curve, left vertical axis) and imaginary (k, blue curve, right vertical axis) parts of the refractive index.

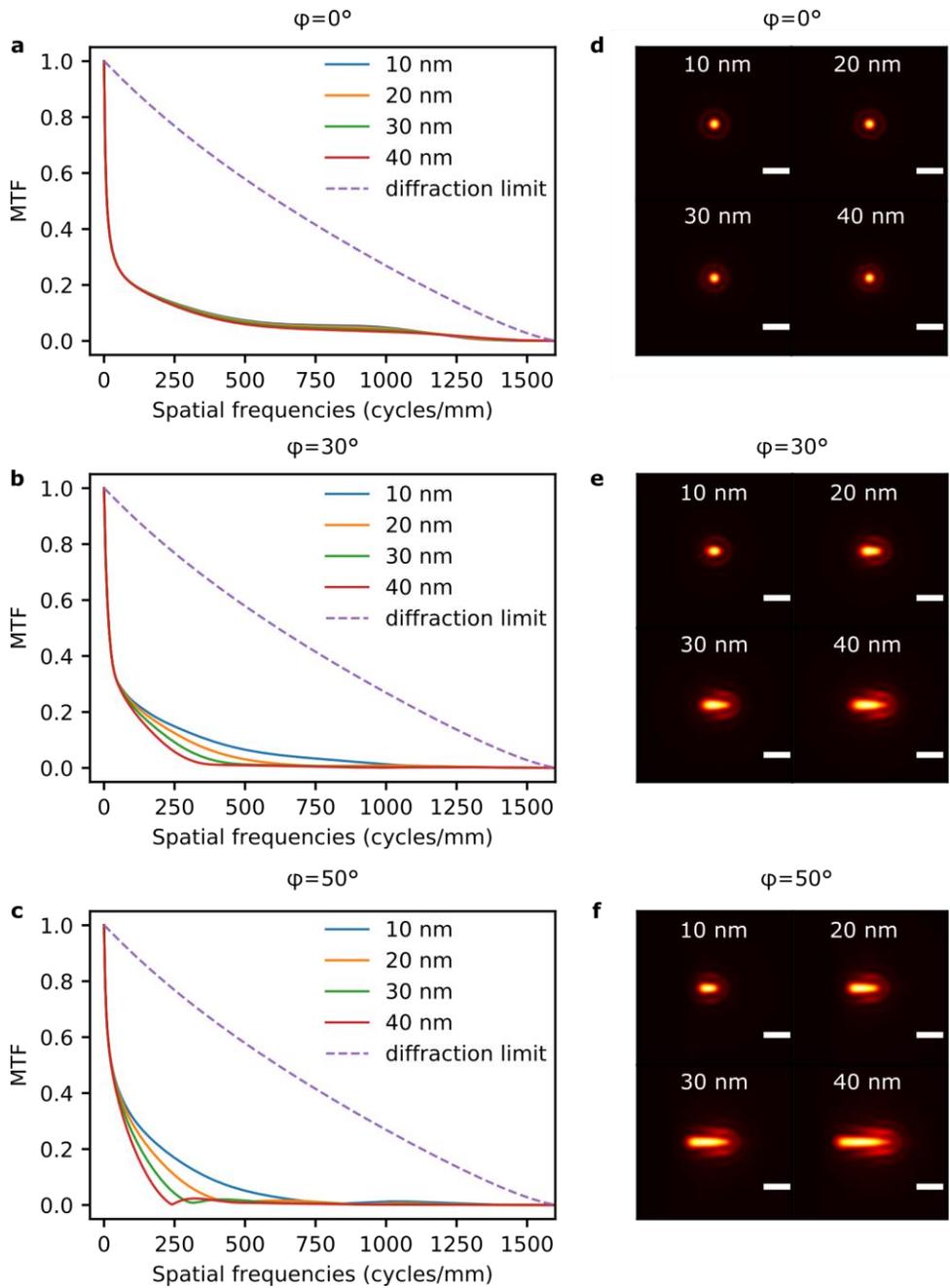

**Figure S2: MTF and PSF simulations of a lens with an ideal quadratic phase profile designed to work in the red (R) region.** (Central wavelength $\lambda_R$ = 620 nm) **a-c** Simulated MTF for angles of incidence $\varphi = 0°, 30°, 50°$ and $\Delta\lambda$ = 10 nm, 20 nm, 30 nm, 40 nm. The diffraction-limited MTF is given for similar $NA$ = 0.5 and $\Delta\lambda$ = 0 nm. d)-f) Simulated PSF for the same conditions. The scale bars correspond to 2 $\mu m$.

# 1 Lateral chromatic aberrations

We demonstrate here the lateral shift difference $\Delta x$ of the PSF position between two different wavelengths $\lambda_1$ and $\lambda_2$. A collimated beam with wavelength $\lambda_1$ impinging a quadratic metalens at an angle of incidence $\varphi$ (in the $xz$ plane, $z$ being the optical axis), in addition to phase delay given by Eq. (4) in the main text, accumulates an extra phase delay $k_1 x \sin(\varphi)$ where $k_1 = 2\pi/\lambda_1$ due to the oblique incidence, compared to a beam coming at normal incidence. Then, the total phase acquired by the beam after the metalens is:

$$\Phi_1(r) = \Phi_1(0) - k_1 \frac{r^2}{2f} + k_1 x \sin(\varphi)$$

$$= \Phi_1(0) - \frac{k_1}{2f}\left[(x - f\sin(\varphi))^2 + y^2\right] + k_1 \frac{f}{2} \sin^2(\varphi) \quad (S.1)$$

where $r = \sqrt{x^2 + y^2}$.

Similarly, a collimated beam with wavelength $\lambda_2$, assuming that this wavelength experiences the same quadratic phase profile as $\lambda_1$ (which means that we neglect material dispersion here), accumulates an extra phase delay $k_2 x \sin(\varphi)$ where $k_2 = 2\pi/\lambda_2$:

$$\Phi_2(r) = \Phi_1(0) - k_1 \frac{r^2}{2f} + k_2 x \sin(\varphi)$$

$$= \Phi_1(0) - \frac{k_1}{2f}\left[\left(x - \frac{k_2}{k_1} f \sin(\varphi)\right)^2 + y^2\right] + \frac{k_2^2}{k_1} \frac{f}{2} \sin^2(\varphi) \quad (S.2)$$

Since in Eqs. (S.1) and (S.2) all the terms independent of the position variables $x$ and $y$ can be omitted (that is the first and last terms in the right hand side of Eqs. (S.1) and (S.2)), these two equations only differ by the terms $x_1 \equiv f \sin(\varphi)$ and $x_2 \equiv \frac{k_2}{k_1} f \sin(\varphi)$, respectively, which correspond to a lateral shift of the PSF in the $x$-direction due to the oblique incidence [1]. Hence, the lateral shift difference given by $\Delta x = x_2 - x_1$ reads:

$$\Delta x = \frac{\Delta k}{k_1} f \sin(\varphi) = -\frac{\Delta \lambda}{\lambda_2} f \sin(\varphi) \quad (S.3)$$

with $\Delta k = k_2 - k_1$ and $\Delta \lambda = \lambda_2 - \lambda_1$.

# 2 MTF analysis

Modulation transfer function (MTF) analysis is a fundamental tool for evaluating imaging performance [2,3]. For this, we measure all R,G and B metalenses using the setup illustrated in Fig. S3. We illuminate them with a collimated laser beam, centred at $\lambda_R$ = 620 nm, 530 nm or 460 nm wavelength, image the point spread function (PSF) and compute its 2D Fourier transform. Since we are interested in large FOV RGB imaging, we conduct the experiment varying the laser angle of incidence ($\varphi$) and bandwidth ($\Delta\lambda$). Fig. S4 presents the measured polychromatic MTF for R(a-c), G (d-f) and B (g-i) channels for $\varphi = 0°, 30°, 50°$ and $\Delta\lambda$ = 10 nm, 20 nm, 30 nm, 40 nm. One can see in Fig. S4 a,d,g that the MTF is almost insensitive to the bandwidth for $\varphi = 0°$. This confirms the robustness of quadratic metalenses against *axial* chromatic aberrations. In Fig. S4b,e,h and c,f,i one can see that MTF starts to slightly degrade for $\varphi = 30°$ and $\varphi = 50°$, following PSF broadening. Experimental MTF (Fig. S4 a,b,c) are in excellent agreement with simulations (Fig. S2a,b,c) for the R channel.

Fig. S5 compares simulated MTFs with and without the aperture stop. Aperture is located at the front focal plane of the lens. The simulations are done for the example of red quadratic metalens ($\lambda_R$ = 620 nm), and with a source bandwidth of 10 nm. MTFs are significantly improved by the aperture stop for both $\varphi = 0°$ and $\varphi = 30°$ cases.

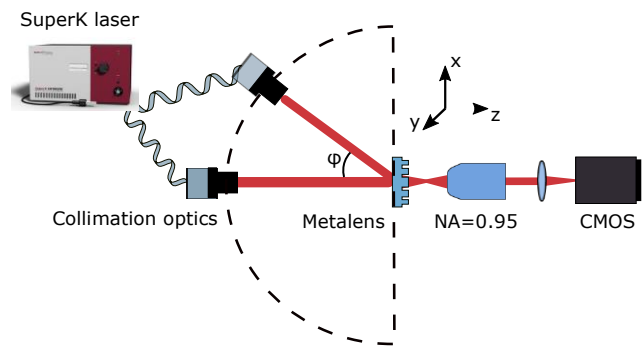

**Figure S3: The optical setup for PSF and MTF characterization.** The metalens is illuminated with a collimated tunable laser source (supercontinuum fiber laser SuperK EXTREME equipped with a tunable single line filter SuperK VARIA), which is placed on a rotation arm. The produced PSF is transferred to a CMOS camera (CS895MU Thorlabs) by a homemade optical microscope (100× Olympus plan apo objective with NA = 0.95 and a tube lens with 150 mm focal length).

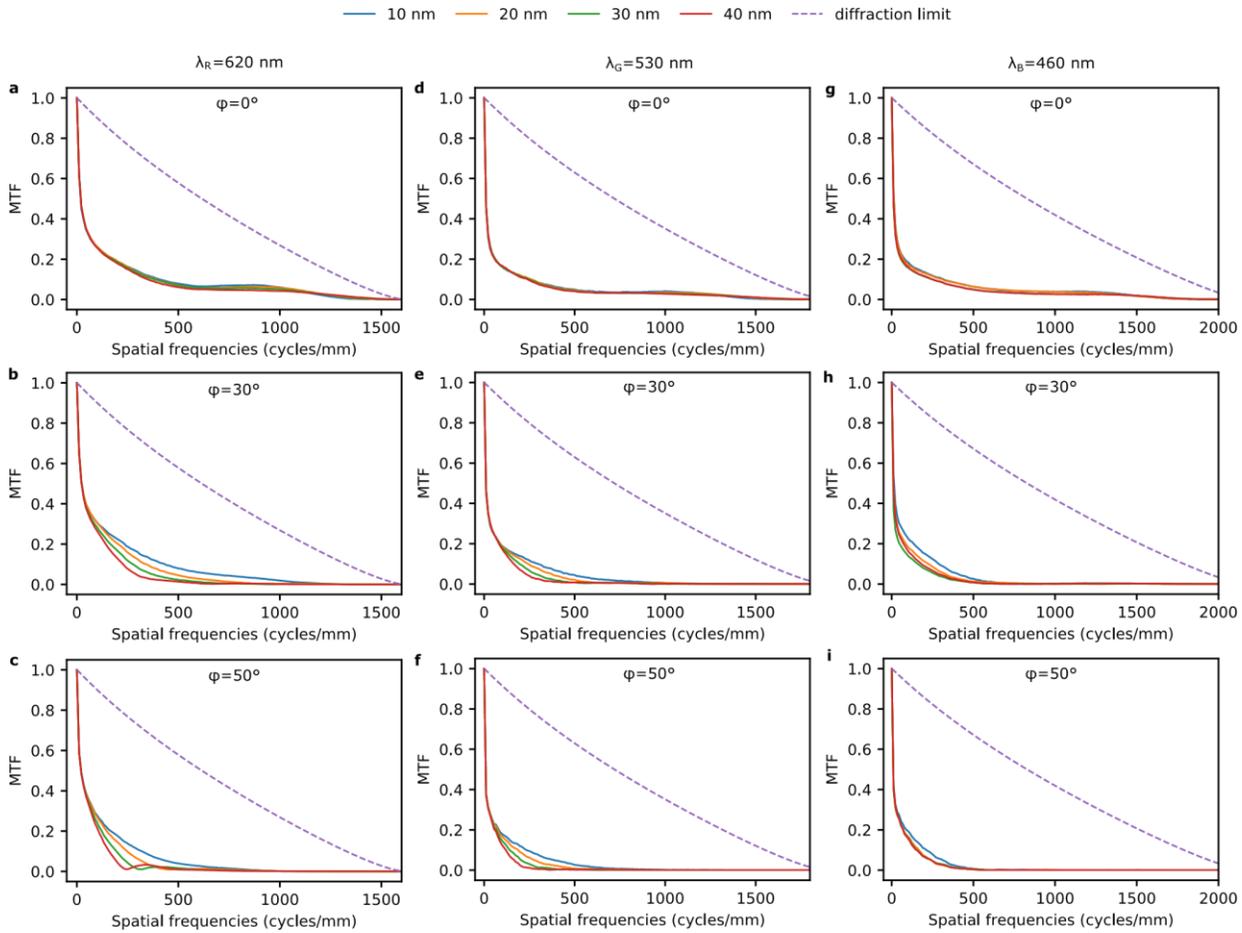

**Figure S4: MTF analysis of the quadratic metalenses in R, G and B channels. a-c** Measured polychromatic MTF for R metalens for incident angles $\varphi$ and source bandwidths $\Delta\lambda$ = 10 nm, 20 nm, 30 nm, 40 nm. **d-f, g-i** Measured polychromatic MTF for G and B metalens correspondingly. The diffraction-limited MTF is given for similar *NA* = 0.48 and $\Delta\lambda$ = 0 nm.

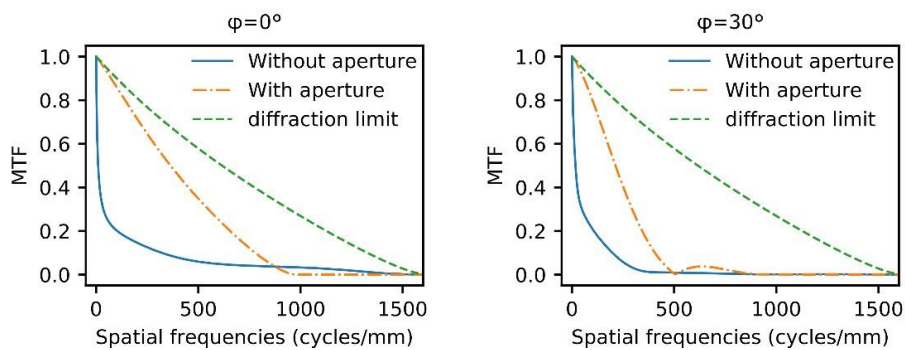

**Figure S5: MTF with aperture stop.** Comparison between simulated MTFs with and without the aperture stop. The simulations are done for the example of red quadratic metalens ($\lambda_R$ = 620 nm), and with a source bandwidth of 10 nm.

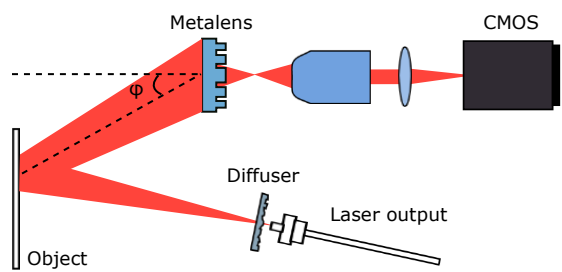

**Figure S6: USAF 1951 target imaging with the quadratic metalens in R channel.** The laser output is diffused to generate spatially incoherent light. The image produced by the metalens was transferred to a CMOS camera (Thorlabs CS895MU) by a homemade optical microscope (100× Olympus plan apo objective with NA = 0.95 and a tube lens with 50 mm focal length)

# 3 Correlation of USAF 1951 imaging with MTF measurements

The studied target element (number 2, group -2) has the spatial frequency of 0.28 cycles/mm in the object plane. Produced element image is squeezed in virtue of the R metalens demagnification. Importantly, the demagnification depends on the angle of view $\varphi$ due to the barrel distortions, leading to a larger compression towards higher $\varphi$. The distance $x$ between the object point and the optical axis is related to the angle of view $\varphi$ as $x = d \tan \varphi$. Since the object is placed far from the metalens (at a distance $d$), light coming from the object point can be approximated as a plane wave. Then, the metalens produces the image of this point located at $x' = f \sin \varphi$ which is a fundamental property of the quadratic phase profile. Next, we make a Taylor expansion of $x'$ and $x$ around $\varphi$, validated by the small angular spread ($\Delta\varphi < 4.5°$ or 0.08 rad). The increments $dx'$ and $dx$ depend of the angular increment $d\varphi$ as:

$$dx \approx d \cdot \frac{d\varphi}{\cos^2 \varphi}; \qquad (S.4)$$

$$dx' \approx f \cdot \cos \varphi \, d\varphi; \qquad (S.5)$$

From these equations we obtain the relation between $dx'$ and $dx$, meaning the metalens demagnification:

$$\frac{dx'}{dx} \approx \frac{f}{d} \cdot \cos^3 \varphi; \qquad (S.6)$$

Note that for $\varphi=0°$ the demagnification is equal to $f/d$, which is a conventional expression for paraxial approximation. Finally, considering that spatial frequencies in the image and object planes are inversely proportional to $dx'$ and $dx$, we calculate that object plane 0.28 cycles/mm translates to ~159, ~244 and ~602 cycles/mm for $\varphi$ = 0°, 30°, 50°, respectively.

To quantify the contrast in each of the images of Fig.3c of the main text, we calculate the contrast transfer function (CTF), which is the function that describes the modulation of a square wave grating in dependence of frequency. These values are shown in Fig. S7 for all resolved cases, together with the corresponding MTF values, calculated using the Coltman formula, which relates the CTF to the MTF [4]. One can see that at normal incidence ($\varphi$ = 0°), the contrast is almost insensitive to the bandwidth, while for oblique incidence ($\varphi$ = 30°), the contrast decreases as the bandwidth increases, which corroborates the observation made on the PSF broadening and MTF quality.

Having calculated the spatial frequencies, we can correlate measured MTF (Figures 2a-2c of the main text) to that extracted from the target element imaging (Figure 3c of the main text). Note that one should take corresponding MTF values at ~159 for $\varphi$ = 0° (Figure 2a), at ~244 for $\varphi$ = 30° (Figure 2b) and at ~602 for $\varphi$ = 50° (Figure 2c). Table 1 summarizes the comparison. One can clearly see a good match, though the values extracted from the imaging (blue color) are slightly elevated. We attribute this to uncertainties in MTF measurements and inaccuracy in the focal plane determination.

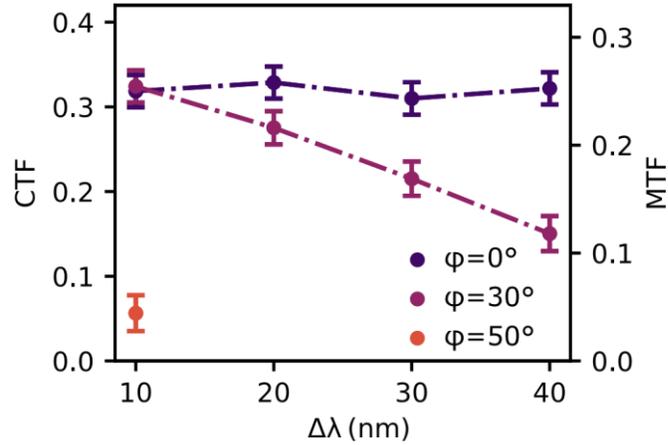

**Figure S7: Contrast Transfer Function.** CTF extracted from imaging with quadratic metalens of the element 2, group -2, R channel (main text Fig. 3c) together with MTF calculated from CTF by Coltman formula.

|  | $\Delta\lambda$ = 10 nm | $\Delta\lambda$ = 20 nm | $\Delta\lambda$ = 30 nm | $\Delta\lambda$ = 40 nm |
|---|---|---|---|---|
| $\varphi = 0°$ | 22/25 | 22/26 | 22/24 | 21/25 |
| $\varphi = 30°$ | 20/25 | 17/21 | 13/17 | 9/12 |
| $\varphi = 50°$ | 3/4 | -/- | -/- | -/- |

**Table S1: The comparison between the MTF measurement done by PSF (black color) and USAF 1951 target imaging (blue color)**

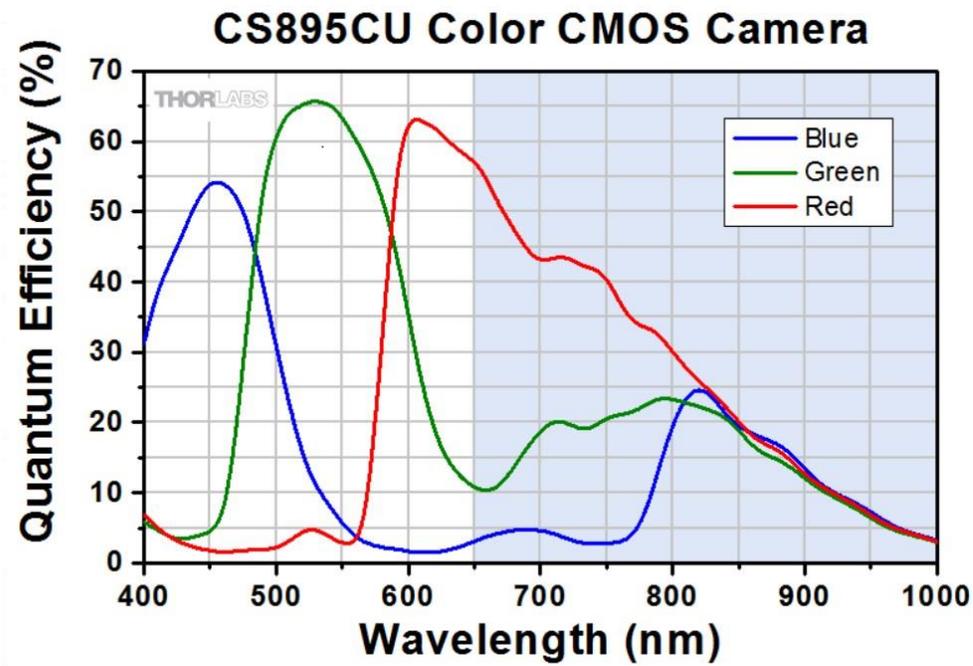

**Figure S8: Camera RGB color filters.** Relative response for the color camera sensor's red, green and blue pixels. The shaded grey region above 650 nm represents wavelengths blocked by the filter. The camera model is Kiralux 8.9 MP CMOS Compact Scientific Cameras from Thorlabs. Image used with the permission from Thorlabs.

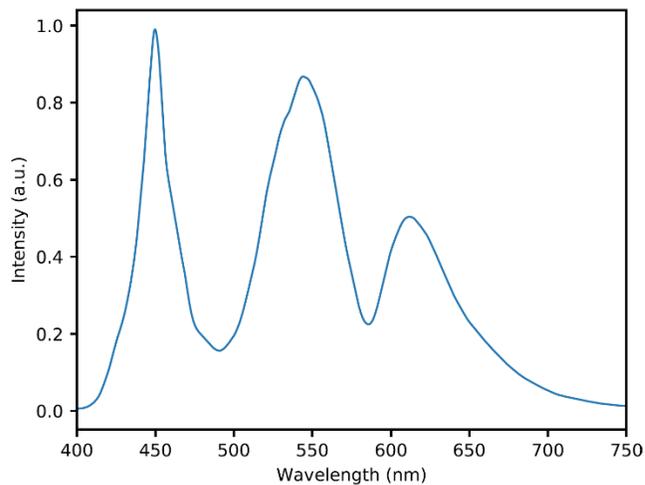

**Figure S9: Xiaomi Redmi Note 8 Pro white color emission spectrum.**

# 4 Color error calculations

To quantify the color reproduction, we make use of the CIELAB metric, which is used to determine the color error based on human vision perception [5]. In this case, RGB intensity values for both reference and measured images are converted to luminance ($L^*$), color relation in red-green ($a^*$) and color relation in yellow-blue ($b^*$). Then, the color error $\Delta E^*$ is calculated as geometric distances in the *L\*a\*b\** three-dimensional space according to:

$$\Delta E^* = \sqrt{(\Delta L^*)^2 + (\Delta a^*)^2 + (\Delta b^*)^2} \qquad (S.7)$$

For FOV of 30° x 20° the color error $\Delta E^*$ (presented in Fig. S10a) is found to be in the range between 5 and 23 and varied for different patches. Panel b and c of Fig. S10 show color errors for 100° x 67° FOV raw image and with intensity correction. To implement the intensity correction procedure in each R, G and B channel, we characterize the efficiency of the quadratic metalenses and use the result as a calibration curve.

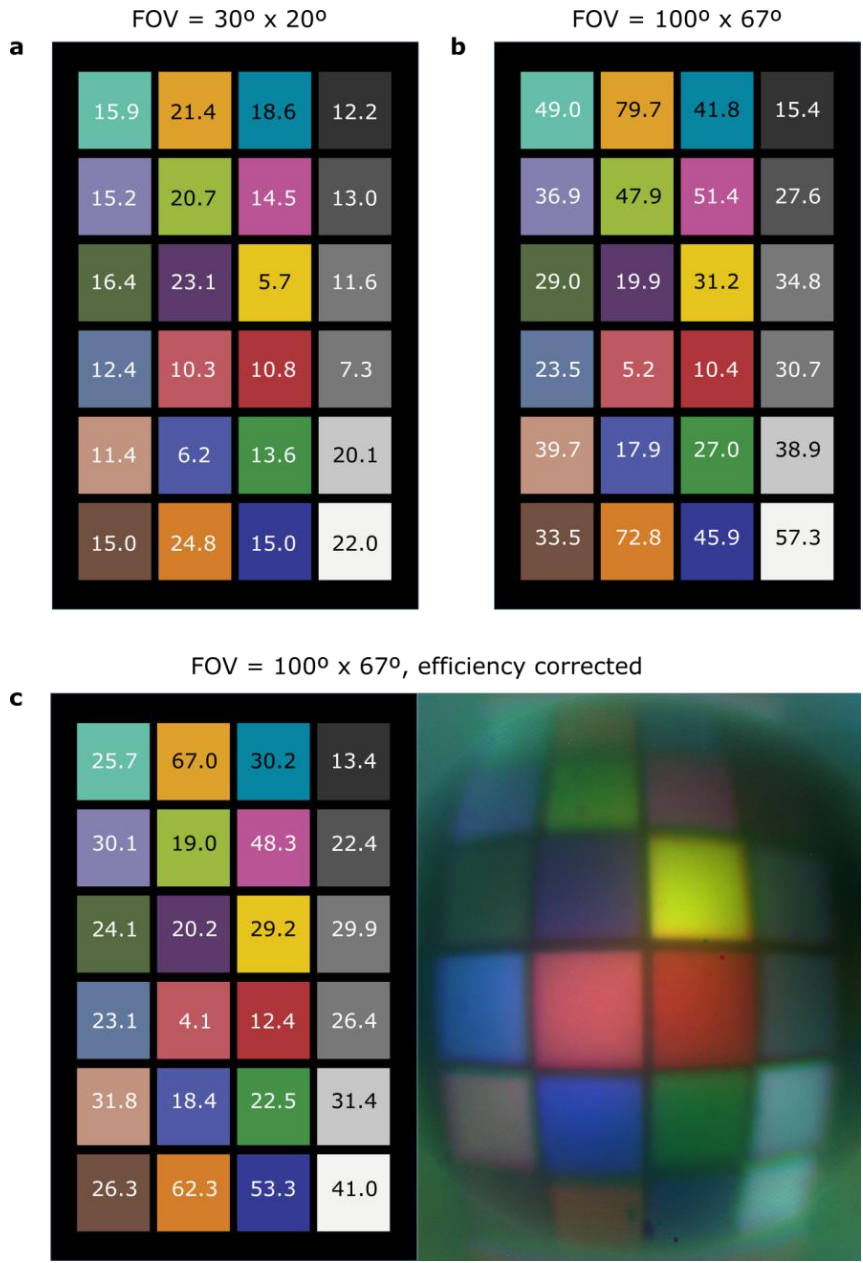

**Figure S10: CIELAB color reproduction assessment. a** The color error Δ$E^*$ for the ColorChecker RGB imaging with FOV of 30° x 20° and **b** 100° x 67°. Δ$E^*$ is given as geometric difference in *L\*a\*b\** three-dimensional space: $\Delta E^* = \sqrt{(\Delta L^*)^2 + (\Delta a^*)^2 + (\Delta b^*)^2}$. **c** The color error (left panel) and obtained RGB image (right panel) after the intensity correction procedure for FOV of 100° x 67°.

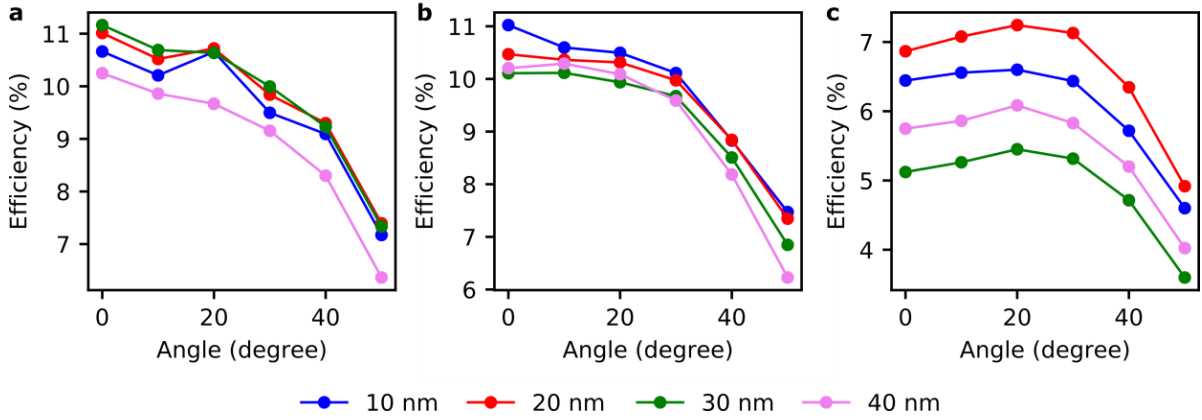

**Figure S11: The focusing efficiency characterization.** Angular dependence of the focusing efficiency at $\Delta\lambda$ = 10 nm, 20 nm, 30 nm, 40 nm for **a** R metalens, **b** G metalens and **c** B metalens.

# 5 EigenCWD Spatially-Varying Deconvolution

The spatially-varying deconvolution algorithm used in our paper seeks a solution, **u**, which minimizes the following objective function [6]:

$$\min_{u} \frac{\mu}{2} \|\boldsymbol{BMu} - \boldsymbol{g}\|_2^2 + \alpha \|\boldsymbol{Cu}\|_1, \qquad (S.8)$$

where **g** is the input blurred image, **BM** is a matrix that describes the spatially varying blur with some decomposition into **B** and **M**, **C** is the total variation operator, **u** is the deblurred estimate of the image, and $\mu$ and $\alpha$ are hyper-parameters. In the original paper [6], the authors developed a column-wise decomposition (CWD) approach in which **B** and **M** were constructed by performing singular-value decomposition (SVD) on the large matrix $\mathbf{H} \equiv \mathbf{BM}$ of size $M^2 \times N^2$, where $M \times N$ is the size of **g**. We refer to this method as the "CWD-SVD deconvolution" in the following. However, performing the matrix multiplication of **Hu**, which is equivalent to evaluating the discretized Rayleigh-Sommerfeld integral, is computationally inefficient. This results in exponentially long runtimes and large memory usage for the images we are interested in deblurring in this paper.

Instead of computing the off-axis PSF from each pixel location in **u** to populate the matrix **H**, as what is done in [6], we implemented the method called "eigenPSF decomposition" described in [7–9] which only requires a small sample of off-axis PSFs to construct a PSF basis. As a result, any off-axis PSF from an arbitrary location within the original image can be approximated by a weighted, linear sum of the PSF basis components. The PSF components, termed as eigenPSFs, are captured in **B**, and the component weights for each pixel location, termed as eigencoefficients, are contained in **M**. The eigenPSF formalism allows us to modify and improve the traditional CWD-SVD deconvolution through more efficient computation especially for larger image sizes. Also, it provides a smooth interpolation of all off-axis PSFs for all spatial points in the image from just a small sample of PSFs.

We proposed and developed an algorithm, that we call the "EigenCWD algorithm", which is the amalgamation of these two methods - the CWD-SVD deconvolution and the eigenPSF decomposition, that enables the spatially-varying deconvolution of the image demonstrated in this article.

The algorithm consists in the following steps:

(i) One first generates a grid of sampled PSFs, such as the 7×5 grid shown in Fig. S12 for example. (ii) The eigenPSFs and eigencoefficients are then computed in a similar fashion as in the eigenPSF decomposition method.

(iii) Next, the computed eigenPSFs and eigencoefficients, as well as the measured image to be deblurred, are sent as inputs into the EigenCWD algorithm to output a deblurred image. For example, in the case treated in this paper (Fig. 5 in the main text), the algorithm requires about 150 iterations to converge.

(iv) Further fine-tuning can be done by using a finer PSF grid to compute a larger basis of eigenPSFs and eigencoefficients, and reiterating the EigenCWD algorithm on the previously obtained output. For example, the reconstruction in Fig. 5 in the main text is the result after another 150 iterations using the 11 × 7 grid.

For the image reconstruction (Fig. 5 in the main text), the point spread functions (PSF) of the R, G and B metalenses were simulated on a 3001 × 3001 simulation grid, with pixel pitch of 82 nm (which corresponds to the original camera's pixel pitch of 3.45 $\mu$m magnified by the 42x objective microscope). For each sampled point on the object, a monochromatic spherical wave was propagated towards the lens, modulated by the lens function, and then propagated a distance of 77 $\mu$m (78 $\mu$m for blue channel) to match our imaging plane distance from the lens (note that these numbers are slightly smaller than the focal distance 83 $\mu$m, and are chosen to correspond to the maximum intensity position, which does not exactly coincide with the focal distance due to the spherical aberrations of the metalens — we made a similar observation in [10]). To create broadband PSFs (as it is the case in the experiment), we generated several monochromatic PSFs across a bandwidth of 40 nm with an interval of 2 nm, centered around the color's channel wavelength. Lastly, since the quadratic metalenses impart a barrel distortion onto the images formed, we corrected for the barrel distortions in both measured images as well as the simulated PSFs before performing the EigenCWD deconvolution. Overall calculation times for full-color image aren't scaled only by input image resolution but implicitly depend on other factors such as PSF grid size and finesse. For example, 1600 x 1200 image takes ∼ 1$h$, while 3400 x 2500 image ∼ 11$h$. Workstation specifications: 28 CPU cores, 2 x E5-2690 v4 @ 2.60GHz; 512GB RAM; 4 GTX TITAN 12GB VRAM.

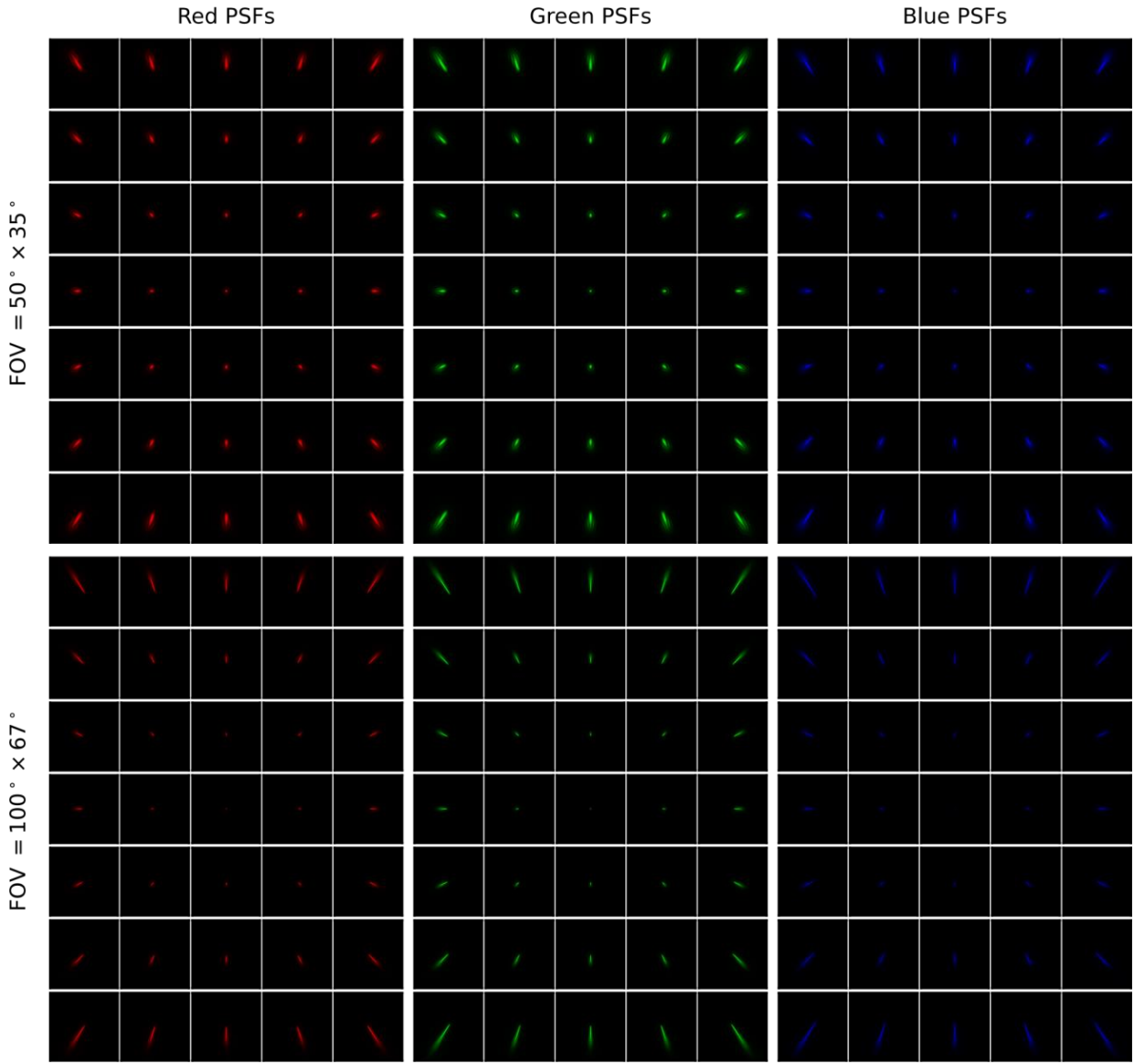

**Figure S12: Sampled broadband PSFs for red (620 nm), green (530 nm) and blue (460 nm) for (top row) 50°× 35° FOV and (bottom row) 100°× 67° FOV**. The barrel distortion has already been removed for the PSFs shown. The 7×5 grids for each color channel implies an even sampling of 7 and 5 points along the vertical and horizontal span of the object respectively. For example, the top left corner of each 7 × 5 represents the PSF of a point source at the top left corner of the object to be imaged at the specified FOV.

# 6 Dispersion-engineering method for scaling-up the system

The metalens bandwidth limits derived in Ref. [11] were obtained for a metalens incorporating a hyperbolic phase profile. In the case of a quadratic phase profile as the one used in this work, following similar derivation as in Ref. [11] leads to the following metalens bandwidth limit:

$$\Delta\omega \leq \frac{2\kappa c}{f} \frac{(1-\text{NA}^2)}{\text{NA}^2} \quad (S.9)$$

where $\kappa$ is given by the Tucker's limit in the case of nanopillars acting as waveguides (used to impart locally a certain phase delay via waveguiding) [11]:

$$\kappa = \frac{\omega_c}{c} H(n_{\max} - n_b) \quad (S.10)$$

with $\omega_c$ the central frequency of the bandwidth $\Delta\omega$, $H$ the height of the nanopillars, and $n_{\max}$ and $n_b$ the refractive indices of the nanopillars and of the background medium, respectively.

By combining Eqs. (S.9) and (S.10), one obtains:

$$\frac{\Delta\omega}{\omega_c} \leq \frac{2H}{f}(n_{\max} - n_b)\frac{(1-\text{NA}^2)}{\text{NA}^2} \quad (S.11)$$

For example, in the case of our red metalens, application of Eq. (S.11) with $D$ = 200$\mu$m, $f$ = 83$\mu$m, (i.e. NA = 0.77), $H$ = 300nm, $n_{\max}$ = 3.3 and $n_b$ = 1 gives a normalized bandwidth of $\Delta\omega/\omega_c$ = 0.0114 which translates into a bandwidth of $\Delta\lambda$ = 7nm around the central wavelength $\lambda_c$ = 620nm (using the fact that $\Delta\lambda/\lambda_c = \Delta\omega/\omega_c$).

Our experimental measurements of more than $\Delta\lambda$ = 40nm bandwidth exceeds this theoretical bandwidth limit but one must remember that the bandwidth limit derivation above assumes a diffraction-limited lens with no aberrations [11], which is not strictly the case for quadratic metalenses because of their intrinsic spherical aberrations.

Nevertheless, one can see from Eq. (S.11) that when scaling-up the size of the metalens while maintaining the same NA, the bandwidth shrinks, as it is inversely proportional to $f$. For example, for a $D$ = 2mm lens with $f$ = 830$\mu$m (corresponding to 10 times the metalenses fabricated in this work), NA is kept constant, but the bandwidth is shrunk by a factor 10 due to $f$, and hence once has to either increase the height of the nanopillars considerably, or the refractive index contrast, or a combination of both, to catch up with the factor of 10. Increasing the height represents a challenge for fabrication, as it leads to higher aspect-ratio, but the progress of fabrication techniques may push the current limits further in the future. Note also that when the height $H$ increases beyond a certain point, the metalens can no longer be considered as an array of one-dimensional delay lines — assumption upon which these bandwidth limits are derived breaks — which may somehow relax the bandwidth limits.